\pdfoutput=1
\documentclass[11pt]{article}

\usepackage{jheppub}
\usepackage{amsmath}
\usepackage{verbatim}
\usepackage{amssymb}
\usepackage{inputenc, array}
\usepackage{textcomp}
\usepackage{bm}
\usepackage{physics}
\usepackage{simpler-wick}
\usepackage{appendix}
\usepackage{xcolor}
\usepackage{comment}
\usepackage{cleveref}
\usepackage{float}
\usepackage{subcaption}
\usepackage{physics}
\usepackage{slashed}
\usepackage{amsfonts,dsfont}
\usepackage{caption}
\usepackage{mathrsfs}
\usepackage{bbm}
\usepackage{graphicx}
\usepackage{standalone}
\usepackage{pgfplots}

\usepackage{hyperref}
 
\newcommand{\be}[1]{\begin{equation}\label{#1} }
\newcommand{\ee}{\end{equation}}
\newcommand{\bes}[1]{\begin{subequations}\label{#1} }
\newcommand{\ees}{\end{subequations}}
\newcommand{\bea}[1]{\begin{eqnarray}\label{#1} }
\newcommand{\eea}{\end{eqnarray}}

\newcommand{\s}{\sigma}

\newcommand{\mt}[1]{\textrm{\tiny #1}}

\newcommand{\eq}[2]{\begin{equation} #1 \label{#2} \end{equation}}

\title{Modular matrices of 2d Carrollian and warped CFTs}

\author[a]{Ankit Aggarwal}
\author[b]{and Joan Sim\'on}

\affiliation[a]{Institute for Theoretical Physics, TU Wien, Wiedner Hauptstrasse~8-10, A-1040 Vienna, Austria} 

\affiliation[b]{School of Mathematics and Maxwell Institute for Mathematical Sciences,\\
University of Edinburgh, Edinburgh EH9 3FD, UK}

\emailAdd{aggarwal@hep.itp.tuwien.ac.at}
\emailAdd{J.Simon@ed.ac.uk}

\abstract{%
We derive the modular S and T matrices that generate the modular transformations on the characters of two dimensional Carrollian and warped CFT algebras. We verify that they satisfy some of their defining  properties.}

\begin{document}
\maketitle

\section{Introduction}

Low-dimensional quantum theories with large amounts of symmetry provide theoretical laboratories for developing our understanding of the world. A prototypical example is 2d conformal field theories (CFTs) where the Virasoro symmetry and modular invariance on the torus have played key roles in condensed matter and holography, but much more broadly in theoretical and mathematical physics   \cite{belavin1984infinite,Ginsparg:1988ui,DiFrancesco:1997nk, Strominger:1997eq}.

In a 2d CFT on a torus with modular parameter $\tau$, modular S-matrices, $\mathbb S(h^\prime,h)$, relate the Virasoro characters $\chi_h(\tau)$ in the direct channel
to the characters in the dual S channel $\chi_{h^\prime}(\tau^\prime=-1/\tau)$  
through
\begin{equation}\label{eq:smatrix}
    \chi_{h^\prime}(\tau^\prime) = \sum_{h} \mathbb S(h^\prime,h)\,\chi_h (\tau)\,,
\end{equation}
with the sum replaced by an integral when the spectrum of $h$ is continuous. Similarly, modular T-matrices relate the Virasoro characters under modular T-transformations $\tau\rightarrow\tau+1$. Together, the S and T transformations generate the full modular group PSL(2,$\mathbb Z$). 
Modular matrices play an important role in the representation theory of Virasoro algebras and theory of 2d CFTs. For instance, modular S-matrices are related to fusion matrices through the Verlinde formula \cite{Verlinde:1988sn}.

 In recent years, two close cousins of 2d CFTs with symmetry algebras not consisting of two Virasoro copies have been studied : 2d Carrollian CFTs (CCFTs) with $\mathfrak{bms}_3$ symmetry algebra (or 2d conformal Carroll algebra) \cite{LevyLeblond1965, Gupta1966} and warped CFTs (WCFTs) with symmetry algebra being one copy of Virasoro in semi-direct product with a $\mathfrak{u}(1)$ Kac-Moody algebra \cite{Hofman:2011zj,Detournay:2012pc}. Carrollian CFTs are expected to play an important role for flat space holography \cite{Bagchi:2010zz,Barnich:2010eb,Bagchi:2012yk,Bagchi:2012xr,Barnich:2012xq,Duval:2014uva,Bagchi:2014iea,Bagchi:2015wna,Hartong:2015usd,Bagchi:2016bcd,Ciambelli:2018wre,Donnay:2022aba,Bagchi:2022emh,Donnay:2022wvx,Bagchi:2023fbj,Saha:2023hsl,Salzer:2023jqv,Saha:2023abr,Mason:2023mti,Chen:2023naw,Nguyen:2023vfz,Alday:2024yyj,Bagchi:2024efs,Bagchi:2024gnn,Ruzziconi:2024kzo}, whereas warped CFTs have been proposed as holographic duals to warped AdS$_3$ and rotating non-extremal and extremal black holes \cite{Compere:2009zj, Detournay:2012pc, Compere:2013bya, Afshar:2015wjm, Apolo:2018oqv,Aggarwal:2019iay, Aggarwal:2020igb, Detournay:2023zni}.

When these theories are put on a torus, they exhibit a modular symmetry that allows one to determine a Cardy-like formula \cite{Cardy:1986ie, Bagchi:2012xr, Bagchi:2019unf, Bagchi:2020rwb, Detournay:2012pc} and universal near-extremal behaviour \cite{Aggarwal:2022xfd,Aggarwal:2023peg,Aggarwal:2025hji}.
There have also been some advances in the representation theory of 2d CCFT and WCFT algebras. For example, the characters are known \cite{Oblak:2015sea, Bagchi:2019unf, Apolo:2018eky}. The purpose of this work is to develop the understanding of the representation-theoretic aspects of these theories/ algebras further. More concretely, we will determine the modular S and T matrices for 2d Carrollian and warped CFT algebras. 

This note is organised as follows. Section \ref{sec:vir-review} revisits the known modular matrices for the Virasoro algebra. Section \ref{sec:strategy} presents the strategy used to find the modular matrices of 2d Carrollian and warped CFTs. Section \ref{sec:carroll} applies this strategy to the 2d conformal Carroll algebra, for both highest weight and induced representations. Section \ref{sec:warped} extends our analysis to the 2d warped CFT algebra, for both non-unitary and unitary representations. In each case we check whether the modular S-matrix squares to the identity or the charge conjugation matrix.

\section{Modular matrices of irrational 2d CFTs/ Virasoro algebra}
\label{sec:vir-review}

Consider a 2d CFT on a torus with modular parameters $\tau, \bar \tau$ that is only invariant under the Virasoro algebra. Since its characters factorise into holomorphic and anti-holomorphic factors, we focus our attention on a single copy of the Virasoro algebra with generators $L_n$. These satisfy the commutation relations
\begin{equation}
    [L_m,L_n]=(m-n)L_{m+n}+\frac{c}{12}n(n^2-1)\delta_{m,-n}
\end{equation}
with $c$ being the central charge.

The modular group acts in the upper half plane, $\rm{Im}~ \tau >0$. It is generated by composing $S$ and $T$ transformations given by
\eq{
    S:\; \tau\to-\frac{1}{\tau}\qquad T:\;\tau\to\tau+1\,.
}{eq:cftmodtrans}
It follows
\eq{
S^2=\mathds{1}~.
}{eq:cft2}

In highest weight representations, primary states are labeled by their weight, $h$,
 \begin{equation}
     L_0|h\rangle=h|h\rangle~,\quad L_{-n}|h\rangle =0~, ~n>0~.
 \end{equation}
Introducing the Liouville momentum parameterization, 
\begin{equation}
    h-\frac{c-1}{24}=:P^2~,\quad \frac{c-1}{6}=:Q^2~, \quad Q=:b+b^{-1}~.
\label{eq:cft-liouville}
\end{equation}
the characters of the Virasoro algebra in such highest weight representation are
\begin{equation} \label{eq:charactersCFT}
 \chi_{\mt{P}}(\tau)=\frac{e^{2\pi i \tau P^2}}{\eta(\tau)}(1-\delta_{\textrm{\tiny vac}} q) ~,\quad q:=e^{2\pi i\tau}\,, \qquad \text{with} \quad
 \delta_{\textrm{\tiny vac}}=\begin{cases}1, & \textrm{for  vacuum}\\
0, & \textrm{otherwise}  \end{cases}\,.
\end{equation}
$\eta(\tau)$ is the Dedekind eta function. Notice the vacuum character is slightly different due to the presence of null states since the 2d CFT vacuum is invariant under the action of $L_{-1}$.

\subsection{Virasoro modular S-matrix.} 
\label{sec:Virasoro}
Consider a generic \textit{non-vacuum} character. The modular S-matrix is defined as the object relating the characters in the direct channel $(\tau)$ to those in the modular S-transformed channel $(\tau^\prime=-1/\tau)$

\begin{equation} \label{eq:CFTSmatrixdef}
    \chi_{\mt{P}'}\left(-\frac{1}\tau\right)=\int_{-\infty}^{\infty} \frac{\dd P}{2}~\mathbb S(P';P) \chi_{\mt{P}}(\tau)\,.
\end{equation}
The $\frac{1}{2}$ factor accounts for the double counting introduced by the quadratic Liouville parameterization \eqref{eq:cft-liouville} of $h$ in terms of $P$.

Non-vacuum characters $\chi_\mt{P}(\tau)$ are gaussian functions of $P$ (see \eqref{eq:charactersCFT}) within the Liouville parameterization \eqref{eq:cft-liouville}. It follows the S-transformed character $\chi_{\mt{P}^\prime}(-\tau^{-1})$ is another gaussian function with an inverse variance dictated by the modular S transformation \eqref{eq:cftmodtrans}. Since Fourier transformations map gaussian functions into gaussian functions with inverse width, as implied by the identity
\begin{equation}
    \int_{-\infty }^{\infty}\dd x \exp\{ax^2\pm bx\}=\sqrt{-\frac{\pi}{a}}\exp \left(-\frac{b^2}{4a}\right)~,\quad \rm{Re  }(a)<0~,
\label{eq:gaussian}
\end{equation}
the Virasoro modular S-matrix for non-vacuum characters should be given by the kernel of a Fourier transformation \cite{Zamolodchikov:2001ah}
\begin{equation}
    \mathbb S(P'; P)=2\sqrt{2}\cos\left[4\pi PP'\right]~.
\label{eq:modular-genCFT}
\end{equation}
Note the modular S-matrix is an even function of $P$ since both $P$ and $-P$ label the same primary state with weight $h$ in the Liouville parameterization \eqref{eq:cft-liouville} . Explicit calculation shows the modular S-matrix \eqref{eq:modular-genCFT} satisfies \eqref{eq:CFTSmatrixdef}
\begin{align}
    &\int_{-\infty}^{\infty} \frac{\dd P}{2} ~ \mathbb S(P';P)\chi_{\mt P}(\tau) =\frac{1}{\eta(\tau)}\frac 1 {\sqrt 2} \sum_{\pm}(\pm)\int_{-\infty}^{\infty}{\dd P}\exp\left[ 2\pi i \left(\tau P^2\pm 2 P P'\right)\right] \cr
&=\frac{1}{\sqrt{-i\tau}\eta(\tau)} \exp\left[ -\frac{2\pi i 
{P'}^2}{\tau}\right] = \chi_{\mt P'}\left(-\frac{1}\tau\right)~,
\end{align}
where \eqref{eq:gaussian} was used with $a=2\pi i\tau$ and $b=4\pi i P'$ \footnote{$\rm{Re  }(a)<0$ since $\rm{Im  }
~\tau > 0$ in the upper half plane.}, together with the modular transformation property of the Dedekind eta function
\begin{equation} \label{eq:Dedekindmod}
    \eta\left(-\frac{1}{\tau}\right)=\sqrt{-i \tau}~\eta(\tau)~.
\end{equation}

 Observe that \eqref{eq:charactersCFT} allows us to write the \textit{vacuum} character $\chi_{\mathds{1}}$ as
 \begin{equation}  
   \chi_{\mathds{1}}=\chi_{\mt P= i\frac{Q}2}-\chi_{\mt P=\sqrt{1-\frac{Q^2}4}} =\chi_{\mt P=\frac{i}{2}(b+b^{-1})}-\chi_{\mt P=\frac{i}{2}(b-b^{-1})}~.
 \end{equation}
 Linearity of \eqref{eq:CFTSmatrixdef} implies the modular S-matrix for the vacuum character satisfies
\begin{equation}
   \mathbb S(\mathds{1};P)= \mathbb S\left(\frac{i}{2}(b+b^{-1});P)\right) - \mathbb S\left(\frac{i}{2}(b-b^{-1});P\right) ~.
\label{eq:cft-linear}
\end{equation}
Using \eqref{eq:cft-linear} together with the identities $\cos x-\cos y=2\sin\frac{x+y}2\sin\frac{y-x}2$ and $\sin(i x)=i\sinh x$, the vacuum modular S-matrix equals\footnote{The same result is obtained by choosing the other branch $P^\prime = -\frac{i}{2}(b\pm b^{-1})$.}
\begin{equation}
   \mathbb S(\mathds{1};P)=4\sqrt{2}\sinh{(2\pi b P)}\sinh{(2\pi b^{-1}P)}~.
\label{eq:Svac-Vir}   
\end{equation}
This is the Plancherel measure for the continuous principal series representations of the quantum group $\mathcal{U}_q(\mathrm{SL}(2,\mathbb{R})$ \cite{Ponsot:1999uf, Ponsot:2000mt}\footnote{$\mathbb S(\mathds{1};P)$ can also be understood as the Plancherel measure on the quantum semi-group $\mathrm{SL}^+_{q}(2,\mathbb{R})$. The latter is an object defined by the spectral decomposition \cite{ip2012representationquantumplanequantum} $\text{L}^{2}(\mathrm{SL}^+_{q}(2,\mathbb{R})) = \int_{\oplus_{p \geq 0} }   \,\mathbb S(\mathds{1};p)\, \mathcal{P}_{p} \otimes \mathcal{P}^{*}_{p}$, 
with $q= e^{\pi i b^{2}}$. This is a quantum group generalization of the Peter-Weyl theorem in which 
$\mathcal{P}_{p}$ are representations of the modular double of $\mathcal{U}_{q}(\mathrm{SL}(2,\mathbb{R}))$. This interpretation allowed for a bulk Hilbert space factorization for 3d gravity with a negative cosmological constant \cite{Mertens:2022ujr}.}. 

The Virasoro modular S-matrix satisfies $ \mathbb S^2=\mathds 1$ \cite{DiFrancesco:1997nk} when acting on highest weight Virasoro characters, i.e.
\begin{align}
 \int_{-\infty}^{\infty} \frac{\dd P'}{2}  \int_{-\infty}^{\infty} \frac{\dd P}{2} ~ \mathbb S(P'';P')\mathbb S(P';P)\chi_{\mt P}(\tau)= \chi_{\mt P''}(\tau)~.
\label{eq:a0}
\end{align}
The integral over $P^\prime$ can be performed exactly
\begin{align}
     \int_{-\infty}^{\infty} \frac{\dd P'}{2} ~\mathbb S(P'';P')\mathbb S(P';P)&=4\int_{-\infty}^{\infty} {\dd P'} ~\cos(4\pi PP')\cos(4\pi PP'')\cr
     &=2\sum_{\pm}\int_{-\infty}^{\infty} {\dd P'} ~\cos(4\pi P'(P\pm P'')) \cr&=\delta(P+P'')+\delta(P-P'') \cr
     &=\delta(|P|-|P''|)
\end{align}
Once again, notice how the quadratic parameterization $P^2$ in the Virasoro characters \eqref{eq:charactersCFT} leads to the discrete symmetry $P \to -P$. Thus, both delta functions above give rise to the same overall character and cancel the extra $\frac{1}{2}$ factor from the remaining $P$ integral in \eqref{eq:a0}.

\subsection{Virasoro modular T-matrix.} The modular T-matrix relates the characters in the direct channel to those in the modular T-transformed channel
\begin{equation} \label{eq:CFTTmatrixdef}
    \chi_{\mt P'}\left(\tau+1\right)=\int_{-\infty}^{\infty} \frac{\dd P}{2}~ \mathbb T(P';P)\chi_{\mt P}(\tau).
\end{equation}
Virasoro characters \eqref{eq:charactersCFT} transform under the T-transformation \eqref{eq:cftmodtrans} due to
\begin{equation}\label{eq:etaTtrans}
    \eta(\tau+1)=e^{\frac{\pi i}{12}}\eta(\tau) \quad \Rightarrow \quad  \chi_{\mt P}(\tau+1)=e^{-\frac{\pi i}{12}}\chi_{\mt P}(\tau)~.
\end{equation}
It follows,
\begin{equation}
    \mathbb T(P';P)=e^{-\frac{\pi i}{12}}\delta(|P|-|P'|)~.
\end{equation}

\section{Strategy to find Carrollian and warped modular S-matrices} 
\label{sec:strategy}

Before deriving the specific modular S-matrices for Carrollian and warped 2d CFTs in Sections \ref{sec:carroll} and \ref{sec:warped}, we present the common strategy motivating our derivation.  Consider either of these theories on a torus with modular parameters $(\tau, z)$ and the S-transformation taking the form
\begin{equation} \label{eq:modSgen}
  S : \tau \to \tau^\prime = -\frac{1}{\tau} \quad \text{and} \quad  z \to z^\prime(\tau,z)~,\quad {\rm Im} (\tau)>0
\end{equation}
where $z'(\tau, z)$ is a known function depending on the theory  (see \eqref{eq:ccft7} or \eqref{eq:wcft7}).  

Our goal is to determine the modular S-matrices for a continuous version of \eqref{eq:smatrix} 
\begin{equation}
    \chi_{\mt P'_\tau,p'}(\tau', z')=\int \dd P_\tau\, \dd p \,\mathbb S(P^\prime_\tau,p^\prime;P_\tau,p)\,\chi_{\mt P_\tau,p}(\tau, z)
\label{eq:cont-Smatrix}
\end{equation}
for characters $\chi_{\mt P_\tau,p}(\tau, z)$ in specific representations of the $\mathfrak{bms}_3$ and the 2d warped algebras. The analogy to the Liouville parameterization \eqref{eq:cft-liouville} will consist of two quantum numbers $(P_\tau,p)$. $P_\tau$ labels the quantum number associated with the observable conjugate to the chemical potential $\tau$ and $p$ labels the quantum number associated with the observable conjugate to $z$. 

Our algorithm will work for characters of the form\footnote{See \eqref{eq:carroll-hw} and \eqref{eq:characters ind} for highest weight and induced 2d Carrollian CFTs, 
\eqref{characters_WCFT_nu} and \eqref{characters_WCFT_u} for non-unitary and unitary 2d warped CFTs.}
\begin{equation} 
  \chi_{\mt P_\tau, p}(\tau,z) = \frac{1}{G(\eta(\tau))}\,\exp[2\pi i\,\tau P_\tau] \exp[2\pi i z s(p)]
\label{eq:c-ansatz}
\end{equation} 
$G(\eta(\tau))$ acknowledges the (known) different functional dependencies in $\eta(\tau)$ accounting for the contribution of descendants in these theories. $s(p)$ is another known function that captures the dependence of the second quantum number $p$.

The integral equation satisfied by the modular S-matrices is obtained by plugging the specific ansatz \eqref{eq:c-ansatz} into \eqref{eq:cont-Smatrix}
\begin{equation}
    \frac{\exp\left[2\pi i \left(-\frac{P^\prime_\tau}{\tau}\right)\right] \exp\left[2\pi i\,z'(\tau,z)\,s(p^\prime)\right]}{G(\sqrt{-i\tau}\,\eta(\tau))} = \int \dd P_\tau\, \dd p \,\mathbb S(P^\prime_\tau,p^\prime;P_\tau,p)\,\frac{\exp\left[2\pi i\tau\,P_\tau\right] \exp\left[2\pi i z\,s(p)\right]}{G(\eta(\tau))}~.
\label{eq:k-problem}
\end{equation}
To solve this, consider the ansatz
\begin{equation}
  \mathbb S(P_\tau^\prime,p^\prime; P_\tau ,p) = {\cal A}(p,p')\,\exp\left[-2\pi i\left(\gamma(p,p^\prime)\,P_\tau + \gamma(p^\prime,p)\,P_\tau^\prime\right)\right]
\label{eq:S-ansatz}
\end{equation}
where $\mathcal{A}(p,p'), \gamma(p,p')$ are arbitrary functions of $p$ and $p'$\footnote{Note this ansatz would not work for standard 2d CFTs.}. The exponential part of the ansatz is
symmetric under $(P_\tau,p)\to (P_\tau ',p')$, a necessary condition for any modular S-matrix\footnote{This is because modular S-matrices are independent of the modular parameters $(\tau,z)$ and modular transformations do not change the spectrum of the theory. It follows the modular S-matrix doesn't see whether it acts on the characters in the direct channel or the dual channel. As a consequence, it must be symmetric.}. Due to the continuous nature of the spectrum in \eqref{eq:cont-Smatrix}, the modular S-matrix ansatz depends on the normalization prefactor $\mathcal A$. The latter is not necessarily symmetric, since it depends on the integration measure. However, in all cases discussed in this note, one can either find a parameterization where it is, or one can split $\mathcal A$ into a symmetric piece and an integration measure. In general, the S-matrix can be a sum of terms of the form \eqref{eq:S-ansatz}, as we will explicitly see for Carrollian CFTs.


When plugging \eqref{eq:S-ansatz} into the right hand side of \eqref{eq:k-problem}, the integral in $P_\tau$ can be performed leading to
\begin{equation}
  \int_{-\infty}^{\infty} \dd P_\tau\,\exp\left[2\pi i\tau\,P_\tau\right]\,\exp\left[-2\pi i\,\gamma(p^\prime,p)\,P_\tau\right] = \delta(\tau - \gamma(p,p^\prime))
\label{eq:Aa}
\end{equation}
The appearance of a delta function in the second momentum charge will localise the remaining integral in \eqref{eq:k-problem}, allowing us to carry it out explicitly. The delta function may localise at more than one value of $p$. In this case, one may need take the ansatz to be a linear superposition of terms of the form \eqref{eq:S-ansatz}, one for each value of $p$ where the delta function localises.

The success of our algorithm relies on the ability to fix the functions $\mathcal A $ and $\gamma$ in \eqref{eq:S-ansatz} by solving \eqref{eq:k-problem} once the above localisation is taken into account. More specifically, the exponentials should agree
\begin{itemize}
\item the remaining exponentials in $P^\prime_\tau$ on either side of \eqref{eq:k-problem} must match
\begin{equation}
  \exp\left[-2\pi i\,\gamma(p^\prime,p_\star)\,P^\prime_\tau\right] = \exp\left[-2\pi i \,\frac{P^\prime_\tau}{\tau}\right] \quad \Rightarrow \quad \gamma(p^\prime,p_\star) = \frac{1}{\tau}\,,
\label{eq:1match}
\end{equation}
where $p_\star$ stands for the solution to $\tau = \gamma(p_\star,p^\prime)$. Hence, altogether we get 
\begin{equation}
   \gamma(p',p_\star) \tau =  \gamma(p',p_\star)\gamma(p_\star,p')=1
\end{equation}
\item exponentials depending on the arbitrary parameterisation $s(p)$ should also match
\begin{equation}
  \exp\left[2\pi i\,z\,s(p_\star)\right] = \exp\left[2\pi i\,z'(\tau,z)\,s(p^\prime)\right] \quad \Rightarrow \quad \frac{s(p_\star)}{s(p')} = \frac{z'(\tau,z)}{z}~.
\label{eq:2match}
\end{equation}
\end{itemize}
Conditions \eqref{eq:1match}-\eqref{eq:2match} must be compatible. The prefactor ${\cal A}$ in the ansatz \eqref{eq:S-ansatz} is fixed in order to deal with the normalisation of the delta function
\begin{equation}
    \delta(\tau-\gamma(p,p^\prime)) = \frac{\delta(p-p_\star)}{|\dd\gamma/\dd p|}
\end{equation}
and the modular S-transformation of the descendant contributions through $G(\eta(\tau))$. This leads to
\begin{equation}
  \frac{1}{G(\sqrt{-i\tau}\,\eta(\tau))} = \frac{{\cal A}}{G(\eta(\tau))}\frac{1}{|\dd\gamma/\dd p|}\,.
\label{eq:A-match}
\end{equation}
In the particular case of $G(\eta(\tau))= (\eta(\tau))^n$\footnote{As we will see, this feature does not hold for non-unitary  warped CFTs with real charge and the methods developed here do not apply to this specific case.}, condition \eqref{eq:A-match} is solved by
\begin{equation}
  {\cal A} = i^{n/2} \left|\frac{\dd\gamma(p,p^\prime)}{\dd p}\right| {\left(\gamma(p,p')\right)^{-n/2}}\,.
\label{eq:A-fix}
\end{equation}

We next show this procedure works for 2d Carrollian CFTs and warped CFTs (unitary and non-unitary with imaginary charges). In particular, we shall describe how the multiple delta function localisation modifies the proposed ansatz \eqref{eq:S-ansatz} by a linear superposition. We further note that if $\gamma(p,p')$ is real $\forall~p,p'$ (as it will turn out to be in the cases we consider), then this procedure would only work for real $\tau$ due to the relation \eqref{eq:1match}.  

\section{Modular matrices of 2d Carrollian CFTs}
\label{sec:carroll}

The symmetry algebra of 2d Carrollian CFTs consists of superrotation generators $L_n$ and supertranslation generators $M_n$ satisfying the commutation relations
\begin{align}
    [L_m,L_n]&=(m-n)L_{m+n}+{c_\mt{L}}~n(n^2-1)\delta_{m,-n}\cr
    [L_m,M_n]&=(m-n)M_{m+n}+{c_\mt{M}}~n(n^2-1)\delta_{m,-n}\cr
     [M_m,M_n]&=0~,
\end{align}
with central extensions $c_\mt{L}$ and $c_\mt{M}$. Alternatively, it can be viewed as the $\mathfrak{bms}_3$ algebra, which is the symmetry algebra of asymptotically flat spacetimes in three dimensions. It can be obtained by In\"onu-Wigner contraction of two copies of Virasoro algebras of standard 2d CFTs, with appropriate central extensions \cite{Bagchi:2010zz}.

Two dimensional Carrollian CFTs can be defined on a torus with modular parameters $(\sigma, \rho)$ \cite{Bagchi:2012xr,Barnich:2012xq,Aggarwal:2025hji}\footnote{See \cite{Aggarwal:2025hji} for a geometric description of this action.} : $\sigma$ being the chemical potential associated with $L_0$ and $\rho$ the chemical potential associated with $M_0$. The corresponding Carroll modular group is generated by composing $S$ and $T$ transformations given by 
\eq{
S:\; \left(\s,\rho\right)\to \left(-\frac{1}{\s},\frac{\rho}{\s^2}\right) \qquad\qquad T:\;\left(\s,\rho\right)\to \left(\s+1,\rho\right)~.
}{eq:ccft7}
These modular transformations have the same form as \eqref{eq:modSgen} and the action of the modular S-transformation satisfies
\eq{
S^2=\mathds{1}~.
}{eq:ccft8}

Carrollian characters are known for highest weight representations \cite{Bagchi:2019unf} and for induced representations \cite{Oblak:2015sea}. We discuss their modular matrices in Sections \ref{sec:h-weight} and \ref{sec:induced}, respectively.

\subsection{Highest weight representations}
\label{sec:h-weight}

Primary states in highest weight representations of the 2d conformal Carroll algebra are labeled by two quantum numbers $(\xi,\Delta)$, corresponding to the eigenvalues of $M_0$ and $L_0$, respectively. Using the two Liouville-like momenta
\begin{equation}
    \xi-\frac{c_\mt{M}}2=:P_\mt{M}^2~,\quad \Delta-\frac{c_\mt{L}-1/6}2=:P_\mt{L}~, \quad c_\mt{L}-\frac{1}{6}=:Q_\mt{L}~,
\label{eq:carroll-par}
\end{equation}
Carrollian characters in highest weight representations can be written as \cite{Bagchi:2019unf}
\begin{equation}
    \chi_{(P_\mt{L},P_\mt{M})}(\sigma, \rho)=\frac{q_\mt{L}^{P_\mt{L}}\,q_\mt{M}^{ P_\mt{M}^2}}{\eta(\sigma)^2}(1-\delta_{\textrm{\tiny vac}}q_\mt{L})^2\,,
\label{eq:carroll-hw}
\end{equation}
where
\begin{equation}
  q_\mt{L}:=e^{2\pi i\sigma}\,, \quad q_\mt{M}:=e^{2\pi i\rho} \quad \text{and} \quad \delta_{\textrm{\tiny vac}}=\begin{cases}1, & \rm{for\; vacuum}\\
0, & \textrm{otherwise}  \end{cases}\,.
\label{eq:qL-qM}
\end{equation}
This character is of the form \eqref{eq:c-ansatz} with the identifications $(\tau, z,P_\tau,p)\rightarrow (\sigma, \rho,P_\mt{L},P_\mt{M})$,   $s(P_\mt{M})= P_\mt{M}^2$, and $G(\eta(\sigma)) = \left(\eta(\sigma)\right)^2$. Therefore, we can follow the analysis of Section \ref{sec:strategy}. We define the Carrollian modular S-matrix to be
\begin{equation} \label{eq:Smatrixdef}
    \chi_{(P_\mt{L}',P_\mt{M}')}\left(-\frac{1}\sigma,\frac \rho{\sigma^2}\right)=\int_{-\infty}^{\infty} \frac{\dd P_\mt{M}}{2} \int_{-\infty}^{\infty}  {\dd P_\mt{L}}~ \mathbb S(P_\mt{L}',P_\mt{M}';P_\mt{L},P_\mt{M})\chi_{(P_\mt{L},P_\mt{M})}(\sigma,\rho)\,,
\end{equation}
where the factor of $2$ takes care of the double counting in $P_\mt{M}$ introduced by our quadratic parameterisation \eqref{eq:carroll-par}. Following \eqref{eq:S-ansatz}, we consider the modular S-matrix ansatz
\begin{equation}
  \mathbb S(P_\mt{L}',P_\mt{M}';P_\mt{L},P_\mt{M}) = {\cal A}(P_\mt{M},P^\prime_\mt{M})\, \exp\left[-2\pi i\left(\gamma(P_\mt{M},P_\mt{M}^\prime)\,P_\mt{L} + \gamma(P_\mt{M}^\prime,P_\mt{M})\,P_\mt{L}^\prime \right)\right]
\label{eq:ansatz-c}
\end{equation}
Since $\pm P_\mt{M}^\prime$ describe the same character, this ansatz must be even under such discrete transformation, i.e. $\mathbb S(P_\mt{L}',-P_\mt{M}';P_\mt{L},P_\mt{M}) = \mathbb S(P_\mt{L}',P_\mt{M}';P_\mt{L},P_\mt{M})$. The same statement holds for $P_\mt{M} \to -P_\mt{M}$.

As in equation \eqref{eq:Aa}, the integral over $P_{\mt L}$ gives rise to the Dirac delta function
\begin{equation}
  \gamma({P_{\mt M}}_\star,P_{\mt M'})=\sigma\,.
\end{equation}
Next, we analyse conditions \eqref{eq:1match}-\eqref{eq:2match}. First, since $s(P_\mt{M}) = P_\mt{M}^2$, equation \eqref{eq:2match} implies 
\begin{equation}
  \left(\frac{{P_\mt{M}}_\star}{P_\mt{M}^\prime}\right)^2 = \frac{1}{\sigma^2}\implies \frac{{P_\mt{M}}_\star}{P_\mt{M}^\prime} =\pm \frac{1}{\sigma}
\end{equation}
Comparing with equation \eqref{eq:1match} suggests the existence of two allowed functions
\begin{equation}
  \gamma_\pm(P_\mt{M},P_\mt{M}^\prime)= \pm\frac{P_\mt{M}^\prime}{P_\mt{M}}~.
\label{eq:gammapm}
\end{equation} 
differing by a sign. Notice these are exchanged under the discrete symmetry $P_\mt{M} \to -P_\mt{M}$, or $P^\prime_\mt{M} \to -P^\prime_\mt{M}$. The corresponding delta function localisations would differ by a sign, but their normalisations would be identical 
\begin{equation}
  \delta \left(\sigma - \gamma_\pm(P_\mt{M},P_\mt{M}^\prime)\right) = \frac{\delta(P_\mt{M}\mp P_{\mt{M}\star})}{|\dd\gamma_\pm/\dd P_\mt{M}|}
\end{equation}
due to the absolute values.

 The above discussion suggests to modify the ansatz \eqref{eq:ansatz-c} to 
\begin{equation}
\begin{aligned}
  \mathbb S(P^\prime_\mt{L},P^\prime_\mt{M};P_\mt{L},P_\mt{M}) &= \sum_\pm{\cal A}_\pm(P_\mt{M},P^\prime_\mt{M}) \exp\left[\pm 2\pi i\,\left(\frac{P^\prime_\mt{M}}{P_\mt{M}}\,P_\mt{L} +  \frac{P_\mt{M}}{P^\prime_\mt{M}}\,P^\prime_\mt{L}\right)\right] ~.
\end{aligned}  
\label{eq:ansatz-c1}
\end{equation}
 Invariance under $P_\mt{M} \to -P_\mt{M}$ requires ${\cal A}_-(P_\mt{M},P_\mt{M}^\prime) ={\cal A}_+(-P_\mt{M},P_\mt{M}^\prime)$ and a similar property under $P^\prime_\mt{M} \to -P^\prime_\mt{M}$. Since both terms in \eqref{eq:ansatz-c1} satisfy constraints \eqref{eq:1match}-\eqref{eq:2match}, one is left to solve the analogous normalisation condition to \eqref{eq:A-match} . This leads to
\begin{equation}
  \frac{i}{\sigma} = \frac{|P_{\mt{M}\star}|^2}{2|P^\prime_\mt{M}|}\left({\cal A}_-(P_{\mt{M}\star},P^\prime_\mt{M}) + {\cal A}_+(-P_{\mt{M}\star},P^\prime_\mt{M})\right) =  \frac{|P_{\mt{M}\star}|^2}{|P^\prime_\mt{M}|}\,{\cal A}_-(P_{\mt{M}\star},P^\prime_\mt{M})
\end{equation}
Using that $\frac{1}{\sigma}=\frac{P^\prime_\mt{M}}{P_{\mt{M}\star}}$, this determines
\begin{equation}
  {\cal A}_-(P_\mt{M},P^\prime_\mt{M}) = i\,\frac{P_\mt{M}}{P^\prime_\mt{M}}\frac{|P^\prime_\mt{M}|}{|P_\mt{M}|^2} \qquad \Rightarrow \qquad
  {\cal A}_+(P_\mt{M},P^\prime_\mt{M}) = -i\,\frac{P_\mt{M}}{P^\prime_\mt{M}}\frac{|P^\prime_\mt{M}|}{|P_\mt{M}|^2}\,.
\end{equation}
To sum up, the modular S-matrix for non-vacuum characters is given by
\begin{equation}
  \boxed{ \mathbb S_{\mathfrak{car}}(P^\prime_\mt{L},P^\prime_\mt{M};P_\mt{L},P_\mt{M})=2\frac {|P^\prime_\mt{M}|} {|P_\mt{M}|^2}\frac{P_\mt{M}}{P^\prime_\mt{M}}\sin\left[2\pi \left(\frac{P^\prime_\mt{M}}{P_\mt{M}}P_\mt{L}+\frac{P_\mt{M}}{P^\prime_\mt{M}}P^\prime_\mt{L}\right)\right]}~.
\label{eq:modular-gen}
\end{equation}
Explicit calculation shows that \eqref{eq:modular-gen} satisfies \eqref{eq:Smatrixdef}. Indeed,
\begin{align}
    &\int_{-\infty}^{\infty} \frac{\dd P_\mt{M}}{2} \int_{-\infty}^{\infty}  {\dd P_\mt{L}}~ \mathbb S(P_\mt{L}',P_\mt{M}';P_\mt{L},P_\mt{M})\chi_{(P_\mt{L},P_\mt{M})}(\sigma,\rho)\cr
&=\sum_{\pm}(\pm)\int_{-\infty}^{\infty}{\dd P_\mt{M}}\frac {|P_\mt{M}'|} {|P_\mt{M}|^2}\frac{P_\mt{M}}{P_\mt{M}'}\frac{e^{2\pi i \rho P_\mt{M} ^2}}{\eta(\sigma)^2}\frac 1 {2i} \int 
    _{-\infty }^{\infty }\dd P_\mt{L}\exp\left[\pm 2\pi i \left(\frac{P_\mt{M}'}{P_\mt{M}}P_\mt{L}+\frac{P_\mt{M}}{P_\mt{M}'}P_\mt{L}'\right)\right]e^{2\pi i \sigma P_\mt{L}}\cr
    &=-i\sum_{\pm}(\pm)\int_{-\infty}^{\infty}{\dd P_\mt{M}}\frac {|P_\mt{M}'|} {|P_\mt{M}|^2}\frac{P_\mt{M}}{P_\mt{M}'}\frac{e^{2\pi i \rho P_\mt{M} ^2}}{\eta(\sigma)^2}\frac 1 2 \exp\left[\pm 2\pi i \frac{P_\mt{M}}{P_\mt{M}'}P_\mt{L}'\right]\delta\left(\sigma\pm\frac{P_\mt{M}'}{P_\mt{M}}\right)\cr
    &=-i\sum_{\pm}(\pm)\int_{-\infty}^{\infty}{\dd P_\mt{M}}\frac {|P_\mt{M}'|} {|P_\mt{M}||\sigma|}\frac{P_\mt{M}}{P_\mt{M}'}\frac{e^{2\pi i \rho P_\mt{M} ^2}}{\eta(\sigma)^2}\frac 1 2 \exp\left[\pm 2\pi i \frac{P_\mt{M}}{P_\mt{M}'}P_\mt{L}'\right]\delta\left(P_\mt{M}\pm\frac{P_\mt{M}'}{\sigma}\right)\cr
   & =\frac{i}{\sigma}\frac{e^{2\pi i \frac{\rho}{\sigma^2} {P_\mt{M}'} ^2}}{\eta(\sigma)^2}e^{-2\pi i \frac{P_\mt{L}'}{\sigma}}\cr
    &=\frac{e^{2\pi i \frac{\rho}{\sigma^2} {P_\mt{M}'} ^2}}{\eta\left(-\frac{1}{\sigma}\right)^2}e^{-2\pi i \frac{P_\mt{L}'}{\sigma}}= \chi_{(P_\mt{L}',P_\mt{M}')}\left(-\frac{1}\sigma,\frac \rho{\sigma^2}\right)~.
\label{eq:one-check}    
\end{align}
We first performed the integral over $P_\mt{L}$ leading to a Dirac delta function localization. Then we used the property $\delta(ax)=\frac{\delta(x)}{|a|}$ and carried out the integral over $P_\mt{M}$. Finally, we used the modular transformation of the Dedekind eta function \eqref{eq:Dedekindmod}.

\paragraph{Vacuum modular S-matrix.} Using \eqref{eq:carroll-hw}, the vacuum character equals
\begin{equation}
        \chi_{\mathds{1}}=\chi_{\left(P_\mt{L}=-\frac{Q_\mt{L}}2, ~P_\mt{M}=i\sqrt{\frac {c_\mt{M}} 2}\right)}-2\chi_{\left(P_\mt{L}=1-\frac{Q_\mt{L}}2, ~P_\mt{M}=i\sqrt{\frac {c_\mt{M}} 2}\right)}+\chi_{\left(P_\mt{L}=2-\frac{Q_\mt{L}}2, ~P_\mt{M}=i\sqrt{\frac {c_\mt{M}} 2}\right)}~.
\end{equation}
By linearity of \eqref{eq:Smatrixdef}, the modular S-matrix for the vacuum character satisfies\footnote{The same expression is obtained choosing the branch $P^\prime_\mt{M} = -i\sqrt{c_\mt{M}/2}$, in agreement with our general arguments regarding the discrete invariance $P^\prime_\mt{M}\to -P^\prime_\mt{M}$.}
\begin{equation}
\begin{aligned}
    \mathbb S(\mathds{1};P_\mt{L},P_\mt{M})&= \mathbb S(-Q_\mt{L}/2,i\sqrt{c_\mt{M}/2};P_\mt{L},P_\mt{M}) - 2\mathbb S(1-Q_\mt{L}/2,i\sqrt{c_\mt{M}/2};P_\mt{L},P_\mt{M}) \\
    &+ \mathbb S(2-Q_\mt{L}/2,i\sqrt{c_\mt{M}/2};P_\mt{L},P_\mt{M})\,.
\end{aligned}
\label{eq:inter-a}
\end{equation}
Using \eqref{eq:modular-gen}, the first contribution in \eqref{eq:inter-a} equals
\begin{equation}
\begin{aligned}
    \mathbb S(-Q_\mt{L}/2,i\sqrt{c_\mt{M}/2};P_\mt{L},P_\mt{M}) &= 2\frac{P_\mt{M}}{|P_\mt{M}|^2} \frac{1}{i} \sin\left[2\pi i\left(\sqrt{\frac{c_\mt{M}}{2}}\frac{P_\mt{L}}{P_\mt{M}} + \frac{P_\mt{M}}{\sqrt{c_\mt{M}/2}}\frac{Q_\mt{L}}{2}\right)\right] \\
    &= 2\frac{P_\mt{M}}{|P_\mt{M}|^2}\,\sinh\left[2\pi\left(\sqrt{\frac{c_\mt{M}}{2}}\frac{P_\mt{L}}{P_\mt{M}} + \frac{P_\mt{M}}{\sqrt{c_\mt{M}/2}}\frac{Q_\mt{L}}{2}\right)\right]
\end{aligned}
\end{equation}
Summing the two remaining contributions in \eqref{eq:inter-a}, we obtain 
\begin{equation}
\begin{aligned}
    \mathbb S(\mathds{1};P_\mt{L},P_\mt{M}) &= 2\frac{P_\mt{M}}{|P_\mt{M}|^2}\left\{\sinh(y+a) - 2\sinh y + \sinh (y-a) \right\} \\
    \text{with}\,\,\,a\equiv-2\pi \frac{P_\mt{M}}{\sqrt{c_\mt{M}/2}} \qquad & \text{and} \qquad y\equiv2\pi\left(\sqrt{\frac{c_\mt{M}}{2}}\frac{P_\mt{L}}{P_\mt{M}} + \frac{P_\mt{M}}{\sqrt{c_\mt{M}/2}}\frac{Q_\mt{L}-2}{2}\right)
\end{aligned}    
\end{equation}
Using $\sinh (y\pm a) = \sinh y\cosh a \pm \cosh y\sinh a$, we observe the $\pm$ contributions cancel, whereas the $\sinh y\cosh a$ contributions add up. Altogether,
\begin{equation}
    \mathbb S(\mathds{1};P_\mt{L},P_\mt{M}) = 4\frac{P_\mt{M}}{|P_\mt{M}|^2}\,\sinh y\,\left(\cosh a - 1\right)\,.
\end{equation}
Using $\cosh 2x - 1=2\sinh^2 x$, the final expression for the vacuum modular S-matrix is
\begin{equation}
    \boxed{\mathbb S_{\mathfrak{car}}(\mathds{1};P_\mt{L},P_\mt{M}) = 8\frac{P_\mt{M}}{|P_\mt{M}|^2}\,\sinh\left[2\pi\left(\sqrt{\frac{c_\mt{M}}{2}}\frac{P_\mt{L}}{P_\mt{M}} + \frac{P_\mt{M}}{\sqrt{c_\mt{M}/2}}\frac{Q_\mt{L}-2}{2}\right)\right]\,\sinh^2  \frac{\pi P_\mt{M}}{\sqrt{c_\mt{M}/2}}}\,.
\end{equation}
Given the interpretation of the Virasoro vacuum modular S-matrix \eqref{eq:Svac-Vir}, one could wonder whether $ \mathbb S_{\mathfrak{car}}(\mathds{1};P_\mt{L},P_\mt{M})$ is a Plancherel measure for continuous representations of some quantum deformation of $\mathrm{ISO}(1,2)$ (see \cite{uqiso3} for one such deformation). A better understanding of the quantum Carrollian Liouville theory \cite{Barnich:2012rz} is required to confirm this speculation. This is left to future work. 

The Carrollian modular S-matrix \eqref{eq:modular-gen} squares to the identity when acting on the characters
\begin{align}
     &\int_{-\infty}^{\infty} \frac{\dd P_\mt{M}'}{2} \int_{-\infty}^{\infty}\dd P_\mt{L}'  ~\mathbb S(P_\mt{L}'',P_\mt{M}'';P_\mt{L}',P_\mt{M}') \left(
     \int_{-\infty}^{\infty} \frac{\dd P_\mt{M}}{2} \int_{-\infty}^{\infty}  {\dd P_\mt{L}} ~\mathbb S(P_\mt{L}',P_\mt{M}';P_\mt{L},P_\mt{M})\chi_{(P_\mt{L},P_\mt{M})}(\sigma,\rho)\right)\cr
     &= \chi_{(P_\mt{L}'',P_\mt{M}'')}(\sigma,\rho)~.
\end{align}
This can be proved by explicit calculation
\begin{align}
     &\int_{-\infty}^{\infty} \frac{\dd P_\mt{M}'}{2} \int_{-\infty}^{\infty}\dd P_\mt{L}' ~\mathbb S(P_\mt{L}'',P_\mt{M}'';P_\mt{L}',P_\mt{M}')\mathbb S(P_\mt{L}',P_\mt{M}';P_\mt{L},P_\mt{M})\cr
     &= 4\int_{-\infty}^{\infty} \frac{\dd P_\mt{M}'}{2} \frac {1} {|P_\mt{M}|^2}\frac {|P_\mt{M}''|} {|P_\mt{M}'|}\frac{P_\mt{M}}{P_\mt{M}''}
   \int_{-\infty}^{\infty}\dd P_\mt{L}' \sin\left[2\pi \left(\frac{P_\mt{M}'}{P_\mt{M}}P_\mt{L}+\frac{P_\mt{M}}{P_\mt{M}'}P_\mt{L}'\right)\right]~\sin\left[2\pi \left(\frac{P_\mt{M}''}{P_\mt{M}'}P_\mt{L}'+\frac{P_\mt{M}'}{P_\mt{M}''}P_\mt{L}''\right)\right]\cr
    &= \int_{-\infty}^{\infty} {\dd P_\mt{M}'}\frac {1} {|P_\mt{M}|^2}\frac {|P_\mt{M}''|} {|P_\mt{M}'|}\frac{P_\mt{M}}{P_\mt{M}''}
   \int_{-\infty}^{\infty}\dd P_\mt{L}'~\sum_{\pm}(\pm) \cos\left[2\pi \left(\left(\frac{P_\mt{L}}{P_\mt{M}}\mp\frac{P_\mt{L}''}{P_\mt{M}''}\right)P_\mt{M}'+\left(\frac{P_\mt{M}\mp P_\mt{M}''}{P_\mt{M}'}\right)P_\mt{L}'\right)\right]\cr
   &= \int_{-\infty}^{\infty} {\dd P_\mt{M}'} \frac {1} {|P_\mt{M}|^2}\frac {|P_\mt{M}''|} {|P_\mt{M}'|}\frac{P_\mt{M}}{P_\mt{M}''}
   \sum_{\pm}(\pm) \cos\left[2\pi \left(\frac{P_\mt{L}}{P_\mt{M}}\mp\frac{P_\mt{L}''}{P_\mt{M}''}\right)P_\mt{M}'\right]\delta\left(\frac{P_\mt{M}\mp P_\mt{M}''}{P_\mt{M}'}\right)\cr
    &= \int_{-\infty}^{\infty} \dd P_\mt{M}'\frac {|P_\mt{M}''|} {|P_\mt{M}|^2}\frac{P_\mt{M}}{P_\mt{M}''}
   \sum_{\pm}(\pm) \cos\left[2\pi \left(\frac{P_\mt{L}}{P_\mt{M}}\mp\frac{P_\mt{L}''}{P_\mt{M}''}\right)P_\mt{M}'\right]\delta\left({P_\mt{M}}\mp{{P_\mt{M}''}}\right)\cr
    &= \delta \left(P_\mt{L}-P_\mt{L}''\right)
\left[
    \delta\left({P_\mt{M}}-{{P_\mt{M}''}}\right)+\delta\left(P_\mt{M}+P_\mt{M}''\right)\right]\cr
    &= \delta \left(P_\mt{L}-P_\mt{L}''\right)
    \delta\left(|P_\mt{M}|-|P_\mt{M}''|\right)
\end{align}
using a similar tools as in \eqref{eq:one-check}.

\paragraph{Carrollian modular T-matrix.} The modular T-matrix relates the characters in the direct channel to those in the modular T-transformed channel
\begin{equation} \label{eq:CCFTTmatrixdef}
    \chi_{(P_\mt{L}',P_\mt{M}')}\left(\sigma+1, \rho\right)=\int_{-\infty}^{\infty} \frac{\dd P_\mt{M}}{2} \int_{-\infty}^{\infty}  {\dd P_\mt{L}}~ \mathbb T(P_\mt{L}',P_\mt{M}';P_\mt{L},P_\mt{M})\chi_{(P_\mt{L},P_\mt{M})}(\sigma,\rho).
\end{equation}
The only non-trivial action of the modular T-transformation \eqref{eq:ccft7} on the Carrollian characters \eqref{eq:carroll-hw} is through the transformation \eqref{eq:etaTtrans} of $\eta(\sigma)$. It follows
\begin{equation}
    \chi_{\mt P_\mt{L},P_\mt{M}}(\sigma+1)=e^{-\frac{\pi i}{6}}\chi_{(P_\mt{L},P_\mt{M})}(\sigma)~.
\end{equation}
Thus,
\begin{equation}
    \mathbb T(P_\mt{L}',P_\mt{M}';P_\mt{L},P_\mt{M})=e^{-\frac{\pi i}{6}}\delta(P_\mt{L}-P_\mt{L}')\delta(|P_\mt{M}|-|P_\mt{M}'|)~.
\end{equation}

\subsection{Induced representations}
\label{sec:induced}

Using the same parameterisation as in \eqref{eq:carroll-par}, Carrollian characters for induced representations \cite{Oblak:2015sea} can be written as
\begin{equation} \label{eq:characters ind}
 \chi^{\rm ind}_{P_\mt{L},P_\mt{M}}(\sigma,\rho)=q_\mt{M}^{P_\mt{M}^2}\frac{q_\mt{L}^{P_\mt{L}}}{|\eta(\sigma)|^2}(1-\delta_{\rm vac}q_\mt{L})^2 ~,
\end{equation}
with $q_\mt{L}$ and and $q_\mt{M}$ as in \eqref{eq:qL-qM}. Notice these equal the characters for the highest weight representations in \eqref{eq:carroll-hw} except for the presence of a modulus in the Dedekind eta function, i.e. $|\eta(\sigma)|$. We briefly discuss the implications of this change for their modular S and T-matrices below.
 
First, consider the modular S-matrix for non-vacuum characters. The entire analysis is analogous to the one for highest representations till the normalization condition step, which now reads
\begin{equation}
  \frac{1}{|\sigma|} =\frac{|P_{\mt{M}*}|^2}{2|P^\prime_\mt{M}|}\left({\cal A}_-(P_{\mt{M}*},P^\prime_\mt{M}) + {\cal A}_+(-P_{\mt{M}*},P^\prime_\mt{M})\right) =  \frac{|P_{\mt{M}*}|^2}{|P^\prime_\mt{M}|}\,{\cal A}_-(P_{\mt{M}*},P^\prime_\mt{M})
\end{equation} 
Since the delta function localization gives rise to $\frac{1}{|\sigma|} = \frac{|P_\mt{M}|}{|P^\prime_\mt{M}|}$, this normalization determines
\begin{equation}
   {\cal A}_- (P_\mt{M},P^\prime_\mt{M})=  {\cal A}_+ (P_\mt{M}, P^\prime_\mt{M}) = \frac{1}{|P_\mt{M}|}\,.
\end{equation}
This leads to the non-vacuum modular S-matrix
\begin{equation}
        \boxed{\mathbb S^{\rm ind}_\mathfrak{car}(P_\mt{L}',P_\mt{M}';P_\mt{L},P_\mt{M})=\frac {2} {|P_\mt{M}|}\cos\left[2\pi \left(\frac{P_\mt{M}'}{P_\mt{M}}P_\mt{L}+\frac{P_\mt{M}}{P_\mt{M}'}P_\mt{L}'\right)\right]}~.
\end{equation}
This is again even in $P_\mt{M}$, as expected, and satisfies both \eqref{eq:Smatrixdef} and $\mathbb (\mathbb S^{\rm ind}_\mathfrak{car})^2=\mathds{1}$.

As for highest weight representations and 2d CFTs, linearity determines the vacuum Carrollian modular S matrix to be
\begin{equation}
    \boxed{\mathbb S^{\rm ind}_\mathfrak{car}(\mathds{1};P_\mt{L},P_\mt{M})=
    \frac {8 } {|P_\mt{M}|}\cosh\left[2\pi \left(\frac{P_\mt{L}}{P_\mt{M}}\sqrt{\frac{c_\mt{M}}{2}}+\frac{P_\mt{M}}{\sqrt{c_\mt{M}/2}} \frac{Q_\mt{L}-2}{2}\right)\right]\sinh^2\left(\frac{\pi P_\mt{M}}{\sqrt{ c_\mt{M}/2}}\right)}~.
\end{equation}

\paragraph{Carrollian modular T-matrix for induced representation.} The absolute value $|\eta(\sigma)|$ appearing in \eqref{eq:characters ind} absorbs the transformation \eqref{eq:etaTtrans} of $\eta(\sigma)$ under modular T-transformations, making Carrollian characters in induced representations \eqref{eq:characters ind} invariant
\begin{equation}
    \chi^{\rm ind}_{P_\mt{L},P_\mt{M}}(\sigma+1)=\chi^{\rm ind}_{(P_\mt{L},P_\mt{M})}(\sigma)~.
\end{equation}
Thus,
\begin{equation}
    \mathbb T_{\rm ind}(P_\mt{L}',P_\mt{M}';P_\mt{L},P_\mt{M})=\delta(P_\mt{L}-P_\mt{L}')\delta(|P_\mt{M}|-|P_\mt{M}'|)~.
\end{equation}


\section{Modular matrices of 2d warped CFTs}
\label{sec:warped}

2d warped CFTs  \cite{Hofman:2011zj, Detournay:2012pc} are non-relativistic quantum field theories with symmetry algebra consisting of the semidirect product of a single Virasoro algebra with a $\mathfrak u (1)$ Kac-Moody algebra. Virasoro generators $L_n$ and $\mathfrak{u}(1)$ Kac-Moody generators $P_n$ satisfy the commutation relations
\begin{align}
    [L_m,L_n]&=(m-n)L_{m+n}+\frac{c}{12}n(n^2-1)\delta_{m,-n}\cr
    [L_m,P_n]&=-nP_{m+n}\cr
     [P_\mt{M},P_n]&=\kappa\frac{n}{2}\delta_{m,-n}~,
\end{align}
 with $\kappa$ the $\mathfrak{u}(1)$ level.
 
Consider a 2d warped CFT on a torus. Given the angular potential $\vartheta$ and the inverse temperature $\beta$, its partition function $Z(\tau,z)$ is written in terms of $\tau = \frac{i\vartheta}{2\pi}$ and $z=\frac{i\beta}{2\pi}$. $\tau$ and $z$ are the chemical potentials associated with $L_0$ and $P_0$ respectively. The modular group for these theories is generated by composing the $S$ and $T$ transformations given by\footnote{It is worth stressing partition functions of  warped CFTs are not invariant under the modular group, but transform as a weak Jacobi form, i.e. $Z(-1/\tau,\beta/\tau) = Z(\tau,\beta)\,\exp\left[-ik \frac{\beta^2}{8\pi\tau}\right]$  \cite{Detournay:2012pc}.}
\eq{
S:\; \left(\tau,z\right) \to \left(-\frac{1}{\tau},\frac{z}{\tau}\right) \qquad\qquad T:\;\left(\tau,z\right) \to\left(\tau+1,z\right)\, .
}{eq:wcft7}
This action on modular parameters satisfies
\eq{
S^2=\mathds{1}~.
}{eq:wcft8}
 
Warped characters are labeled by two charges $(h,p)$ : $h$ is the eigenvalue of the angular momentum $L_0$ and $p$ is the eigenvalue of the hamiltonian $P_0$. The warped CFT algebra has both unitary and non-unitary representations. Non-unitary ones were conjectured to be relevant for holography in \cite{Aggarwal:2022xfd}. The modular matrix analysis below is split into non-unitary and unitary representations.

\subsection{Non-unitary Warped CFTs}

Non-unitary  warped CFTs have negative level, i.e. $\kappa<0$. Their characters depend on whether the charge $p$ is purely real or imaginary\footnote{We don't consider complex charge $p$ because its gravity dual metric become complex and it is not clear what their significance for holography is \cite{Apolo:2018eky}.}. The warped modular S-matrix is the object relating the characters in the direct channel to those in the modular S-transformed channel
\begin{equation} \label{eq:SmatrixdefWCFT}
    \chi_{h',p'}\left(-\frac{1}\tau,\frac z{\tau}\right)=\int_{-\infty}^{\infty} \dd h \int_{-\infty}^{\infty}  {\dd p}~ \mathbb S(h',p';h,p)\chi_{h,p}(\tau, z).
\end{equation}
\paragraph{Real charge.} Characters with real charge were computed in \cite{Apolo:2018eky}. In the notation of \cite{Aggarwal:2022xfd}, these are
\begin{equation}\label{characters_WCFT}
		\chi^{(1)}_{h,p}(\tau,z) = \frac{1}{\eta(2\tau)} q^{h- \frac{c-2}{24}} y^p \left(1-\delta_{\text{vac}}q\right)~, \qquad p\in \mathbb{R}~,
\end{equation}
where,
\begin{equation}
q=e^{2\pi i \tau}, \quad y=e^{2\pi i z}, \quad \delta_{\textrm{\tiny vac}}=\begin{cases}1, & \text{for vacuum}\\
0, & \textrm{otherwise}  \end{cases}\,.
\label{eq:wcft-def}
\end{equation}
Comparing with \eqref{eq:c-ansatz}, we learn $G(\eta(\tau)) = \eta(2\tau)$. 
Under S-transformation $\eta(2\tau)\to\eta(-2/\tau)$. We are not aware of a universal relation between $\eta(2\tau)$ and $\eta(-2/\tau)$, i.e. valid for any value of $\tau$. Hence, the strategy described in Section \ref{sec:strategy} will not hold for this case and we are not able to obtain its modular S-matrix.

\paragraph{Imaginary charge.} Non-vacuum characters for a non-unitary warped CFT for a primary with imaginary charge were computed in \cite{Apolo:2018eky}. In the notation of \cite{Aggarwal:2022xfd}, these equal
\begin{equation}\label{characters_WCFT_nu}
		\chi^{(2)}_{h,p}(\tau,z) = \frac{1}{\eta(\tau)^2} q^{h- \frac{c-2}{24}} y^p \left(1-\delta_{\text{vac}}q\right)~, \qquad p\in i\,\mathbb{R}~.
\end{equation}
with $q$ and $y$ as in \eqref{eq:wcft-def}.

Following the general discussion in Section \ref{sec:strategy}, we introduce two Liouville-like momenta
\begin{equation}
  h - \frac{c-2}{24} =: P_\mt{W}\,, \quad p =: i\,p_\mt{im}\,,
\end{equation}
for the $\tau$ and z-cycles, respectively. The modular S-matrix is then defined as 
\begin{equation} \label{eq:SmatrixdefWCFTim}
    \chi^{(2)}_{P_\mt{W}',p'_\mt{im}}\left(-\frac{1}\tau,\frac z{\tau}\right)=\int_{-\infty}^{\infty} \dd P_\mt{W} \int_{-\infty}^{\infty}  {\dd p}_\mt{im}~ S^{(2)}(P_\mt{W}',p'_\mt{im};P_\mt{W},p_\mt{im})\chi^{(2)}_{P_\mt{W},p_\mt{im}}(\tau, z)~.
\end{equation}
The chosen parameterization corresponds to $s(p_{\rm im}) = p_{\rm im}$ in \eqref{eq:c-ansatz} and leads to the S-matrix ansatz
\begin{equation}
  \mathbb S^{(2)}(P_\mt{W}',p'_\mt{im};P_\mt{W},p_\mt{im}) = {\cal A}\exp\left[-2\pi i\left(\gamma(p_\mt{im},p^\prime_\mt{im})\,P_\mt{W} + \gamma(p^\prime_\mt{im},p_\mt{im})\,P^\prime_\mt{W}\right) \right]\,.
\end{equation}
When performing the $P_\mt{W}$ integral in \eqref{eq:SmatrixdefWCFTim}, it gives rise to $\delta(\tau - \gamma(p_\mt{im},p^\prime_\mt{im}))$. Conditions \eqref{eq:1match}-\eqref{eq:2match} uniquely determine\footnote{When solving \eqref{eq:2match}, we used $z'(\tau,z)=\frac{z}{\tau}$, as given in \eqref{eq:wcft7}.}
\begin{equation}
  \gamma(p_\mt{im},p^\prime_\mt{im}) = \frac{p^\prime_\mt{im}}{p_\mt{im}}\,.
\end{equation}
This guarantees all remaining exponentials in both sides of equation \eqref{eq:SmatrixdefWCFTim} match. Thus, we are only left with the normalisation conditions \eqref{eq:A-match}-
\eqref{eq:A-fix} fixing ${\cal A}$
\begin{equation*}
  \frac{1}{(-i\tau)(\eta(\tau))^2} = \frac{{\cal A}}{(\eta(\tau))^2} \frac{|p^2_\mt{im}|}{p^\prime_\mt{im}} \quad \Rightarrow \quad {\cal A} = \frac{i}{\tau} \frac{|p^\prime_\mt{im}|}{|p^2_\mt{im}|} = i \frac{|p^\prime_\mt{im}|}{|p^2_\mt{im}|} \frac{p_\mt{im}}{p^\prime_\mt{im}}
\end{equation*}
where in the last step we used the fact that this normalisation condition is evaluated on the localisation of the delta function, i.e. $\tau = \frac{p^\prime_\mt{im}}{p_\mt{im}}$.

To sum up, the modular S-matrix for non-vacuum characters is given by 
\begin{equation}
  \boxed {\mathbb S_{\rm wcft}^{(2)}(P_\mt{W}',p'_\mt{im};P_\mt{W},p_\mt{im})=i\,\frac{p_\mt{im}}{p'_\mt{im}}\,\frac{|p'_\mt{im}|}{|p_\mt{im}|^2}\exp\left[-2\pi i
        \left(\frac{p'_\mt{im}}{p_\mt{im}}P_{\mt{W}}+\frac{p_\mt{im}}{p'_\mt{im}}P_{\mt{W}}'\right)\right]}~.
\label{eq:warp-img}
\end{equation}
From \eqref{characters_WCFT_nu}, the vacuum character $\chi_{\mathds{1}}$ satisfies
\begin{equation}
        \chi_{\mathds{1}}(\tau,z)=\chi_{\mt P_{\mt{W}}^{\rm vac}, ~p_\mt{im}^{\rm vac}}(\tau,z)-\chi_{1+P_\mt{W}^{\rm vac}, ~p_\mt{im}^{\rm vac}}(\tau,z)~,
\end{equation}
where $p_\mt{im}^{\rm vac}$ and $P_\mt{W}^{\rm vac}$ are the values of $p_\mt{im}$ and $P_\mt{W}$ for the vacuum state, respectively. 
Linearity determines  the vacuum character with imaginary charge, $p_\mt{im}^{\rm vac}$, to be
\begin{equation} \label{eq:SmatrixvacW}
   \boxed{\mathbb S_{\rm wcft}^{(2)}(\mathds{1};P_\mt{W},p_\mt{im}^{\rm vac})=-\frac {2p_\mt{im}}{p_\mt{im}^{\rm vac}}\frac{|p_\mt{ im}^{\rm vac}|}{|p_\mt{im}|^2}\exp\left[-2\pi i  \left(\frac{p_\mt{im}^{\rm vac}}{p_\mt{im}}P_{\mt{W}}+\frac{p_\mt{im}}{p_\mt{im}^{\rm vac}}\left(P_{\mt W}^{\rm vac}+\frac 1 2\right)\right)\right]\sin\left(\frac{\pi p_\mt{im}}{p_\mt{im}^{\rm vac}}\right)}~.
\end{equation}

\subsection{Unitary Warped CFTs}

Unitary warped CFTs have positive level, i.e. $\kappa>0$, and real charge $p$. Their characters were computed in \cite{Apolo:2018eky}. Using the notation in \cite{Aggarwal:2022xfd}, they are
\begin{equation}\label{characters_WCFT_u}
		\chi^{(3)}_{h,p}(\tau,z) = \frac{1}{\eta(\tau)^2} q^{h- \frac{c-2}{24}} y^p \left(1-\delta_{\text{vac}}q\right)~, \qquad p\in \mathbb{R}~.
\end{equation}
Mathematically, all functional dependences in these characters are as in \eqref{characters_WCFT_nu}, but with $p\in \mathbb{R}$. We can thus use the same parameterization without introducing $p_{\rm im}$ and reaching the same conclusion. The modular S-matrix for non-vacuum characters equals
\begin{equation}
    \boxed{\mathbb S_{\rm wcft}^{(3)}(P_\mt{W}',p';P_\mt{W},p)=i{\frac {p} {p'}}\frac{|p'|}{|p|^2}\exp\left[-2\pi i 
    \left(\frac{p'}{p}P_{\mt{W}}+\frac{p}{p'}P_{\mt{W}}'\right)\right]}~,
\label{eq:S-warped-u}
\end{equation}
whereas for the vacuum character, it is
\begin{equation}
    \boxed{\mathbb S_{\rm wcft}^{(3)}(\mathds{1};P_\mt{W},p)=-\frac {2p} {p_{\rm vac}}\frac{|p_{\rm vac}|}{|p|^2}\exp\left[-2\pi i  \left(\frac{p_{\rm vac}}{p}P_{\mt{W}}+\frac{p}{p_{\rm vac}}\left(P_{\mt W}^{\rm vac}+\frac 1 2\right)\right)\right]\sin\left(\frac{\pi p}{p_{\rm vac}}\right)}~.
\end{equation}

\subsection{Checking a further modular S-matrix property}
Even though the action of the modular group on the modular parameters satisfies $S^2=\mathds{1}$, the same group action on characters is known to differ in the presence of extra symmetries. This is because characters may not provide a faithful representation of the modular group \cite{DiFrancesco:1997nk}. In such situations, the more general condition satisfied by the modular S-matrix action on the characters is
\begin{equation} \label{eq:chargeconj}
  \mathbb S^2 = C\,,
\end{equation}
where $C$ is the charge conjugation matrix, leading to the more general unitarity condition
\begin{equation}
  \mathbb S\,\mathbb S^\dagger = \mathds{1}\,.
\label{eq:ssdagger}
\end{equation}
Due to the presence of the $\mathfrak{u}(1)$ Kac-Moody algebra in the 2d warped algebras discussed above, we next check whether the modular S-matrices $\mathbb S_{\rm wcft}^{(2)}$ and $\mathbb S_{\rm wcft}^{(3)}$ satisfy this requirement.

\paragraph{Non-unitary warped modular S-matrix with imaginary charge.}
In \eqref{eq:warp-img}, the non-vacuum modular S-matrix for these theories was shown to equal
\begin{equation}
    \mathbb S^{(2)}(P_\mt{W}',p'_\mt{im};P_\mt{W},p_\mt{im})=i\,\frac{p_\mt{im}}{p'_\mt{im}}\,\frac{|p'_\mt{im}|}{|p_\mt{im}|^2}\exp\left[-2\pi i
    \left(\frac{p'_\mt{im}}{p_\mt{im}}P_{\mt{W}}+\frac{p_\mt{im}}{p'_\mt{im}}P_{\mt{W}}'\right)\right]~.
\end{equation}
We remind the reader the actual charge is purely imaginary, i.e., $p=ip_\mt{im}$. We want to show condition \eqref{eq:ssdagger} 
\begin{equation}
  \int_{-\infty}^{\infty} \dd P^\prime_\mt{W} \int_{-\infty}^{\infty}\dd p^\prime_\mt{im} ~\mathbb S^{(2)}(P''_\mt{W},p''_\mt{im};P^\prime_\mt{W},p^\prime_\mt{im})\left(\mathbb S^{(2)}(P^\prime_\mt{W},p^\prime_\mt{im};P_\mt{W},p_\mt{im})\right)^\dagger = \delta(P''_\mt{W}-P_\mt{W}) \delta(p''_\mt{im}-p_\mt{im})
\end{equation}
holds. Plugging in the modular S-matrices, the integral over $P'_\mt{W}$ reduces to
\begin{equation}
    \int_{-\infty}^{\infty} \dd P^\prime_\mt{W} \,\exp[-2\pi iP'_\mt{W}\left(\frac{p''_\mt{im}}{p'_\mt{im}}-\frac{p_\mt{im}}{p'_\mt{im}}\right)]  = \delta(p''_\mt{im}-p_\mt{im}) |p'_\mt{im}|
\end{equation}
Inserting this into the remaining integral over $p'_\mt{im}$, the support of the delta function $\delta(p''_\mt{im}-p_\mt{im})$ allows us to write
\begin{equation}
    \frac{1}{|p_\mt{im}|} \int^{\infty}_{-\infty} \dd p'_\mt{im}\,\exp[-2\pi i \frac{p'_\mt{im}}{p_\mt{im}}\left(P''_\mt{W} -P_\mt{W}\right)] \delta(p''_\mt{im}-p_\mt{im}) =  \delta(P''_\mt{W}-P_\mt{W}) \delta(p''_\mt{im}-p_\mt{im})~,
\end{equation}
as required. Here, we used the modular S-matrix is symmetric in $(P_{\mt W},p)\to (P_{\mt W}',p')$ as discussed in Section \ref{sec:strategy}\footnote{To reach this conclusion one views $\frac{\dd p_{\rm im}}{|p_{\rm im}|}$ as part of the integration measure and not of the modular S-matrix. Thus, the $\dagger$ operation does not act on the integration measure.}. Therefore, the action of $\dagger$ only amounts to complex conjugation of the modular S-matrix. 

One can similarly check that
\begin{equation}
  \int_{-\infty}^{\infty} \dd P^\prime_\mt{W} \int_{-\infty}^{\infty}\dd p^\prime_\mt{im} ~\mathbb S^{(2)}(P''_\mt{W},p''_\mt{im};P^\prime_\mt{W},p^\prime_\mt{im})\mathbb S^{(2)}(P^\prime_\mt{W},p^\prime_\mt{im};P_\mt{W},p_\mt{im}) = \delta(P''_\mt{W}-P_\mt{W}) \delta(p''_\mt{im}+p_\mt{im})~.
\end{equation}
This implies $\mathbb (\mathbb S^{(2)}_{\rm wcft})^2$ takes the charge $p\rightarrow -p$ as expected from \eqref{eq:chargeconj}.
\paragraph{Unitary warped modular S-matrix with real charge.} Due to the similarity between the modular S-matrix in this case (see equation \eqref{eq:S-warped-u}) and the one in \eqref{eq:warp-img}, the last calculation shows
\begin{equation}
  \int_{-\infty}^{\infty} \dd P^\prime_\mt{W} \int_{\infty}^{\infty}\dd p^\prime ~\mathbb S^{(3)}(P''_\mt{W},p'';P^\prime_\mt{W},p^\prime)\left(\mathbb S^{(3)}(P^\prime_\mt{W},p^\prime;P_\mt{W},p)\right)^\dagger = \delta(P''_\mt{W}-P_\mt{W}) \delta(p''-p)~
\end{equation}
and
\begin{equation}
  \int_{-\infty}^{\infty} \dd P^\prime_\mt{W} \int_{-\infty}^{\infty}\dd p^\prime ~\mathbb S^{(3)}(P''_\mt{W},p'';P^\prime_\mt{W},p^\prime)\mathbb S^{(3)}(P^\prime_\mt{W},p^\prime;P_\mt{W},p) = \delta(P''_\mt{W}-P_\mt{W}) \delta(p''+p)~.
\end{equation}
Hence, the modular S-matrices for unitary warped  CFTs in \eqref{eq:S-warped-u} do satisfy \eqref{eq:ssdagger} and \eqref{eq:chargeconj}.

\subsection{Warped modular T-matrices}

We end this discussion by computing the modular T-matrices relating the characters in the direct channel to those in the modular T-transformed channel
\begin{equation} \label{eq:WCFTmatrixdef}
    \chi_{h',p'}\left(\tau+1,z\right)=\int_{-\infty}^{\infty} {\dd h}~ \int_{-\infty}^{\infty} {\dd p}~ \mathbb T(p',h';p,h)\chi_{h,p}(\tau,z)\,.
\end{equation}
As in previous discussions, all warped characters \eqref{characters_WCFT}, \eqref{characters_WCFT_nu}, and \eqref{characters_WCFT_u} only transform through the Dedekind $\eta(\tau)$ transformation in \eqref{eq:etaTtrans}
\begin{equation}
    \chi^{(1)}_{h,p}(\tau+1,z)=e^{-\frac{\pi i}{12}}
    \chi^{(1)}_{h,p}(\tau,z)~,\quad \chi^{(2,3)}_{h,p}(\tau+1,z)=e^{-\frac{\pi i}{6}}
    \chi^{(2,3)}_{h,p}(\tau,z)~,
\end{equation}
It follows
\begin{equation}
    \mathbb T^{(1)}(h',p';h,p)=e^{-\frac{\pi i}{12}}\delta(h-h')\delta(p-p')~,
    \end{equation}
    \begin{equation}
    \mathbb T^{(2,3)}(h',p';h,p)=e^{-\frac{\pi i}{6}}\delta(h-h')\delta(p-p')~.
    \end{equation}
    
\section{Discussion}

This work presents some new results regarding modular S and T matrices of Carrollian and warped 2d CFTs for different representations.
We first rederived the known modular S-matrices for irrational 2d CFTs in Subsection \ref{sec:Virasoro} using a well-known property of Gaussian distributions \eqref{eq:gaussian}. In Section \ref{sec:strategy}, an strategy was developed to compute modular S-matrices for 2d Carrollian and warped CFTs, with characters of the form \eqref{eq:c-ansatz}. Such strategy uses a convenient Liouville-like parameterization for the quantum numbers associated with the relevant observables and does not rely on any factorization properties of the characters, in sharp contrast with the standard 2d CFT case. The functional form of the characters and the modular S-matrix gives rise to a delta function localisation that trivialises the integral over the second quantum number. This strategy was applied to unitary warped CFTs and non-unitary warped CFTs with imaginary charge in Section \ref{sec:warped}. When the delta function localisation equation has more than one root, such as for 2d Carrollian CFTs, the original ansatz \eqref{eq:c-ansatz} is modified to a linear superposition, as in \eqref{eq:ansatz-c1}. We have shown that this works for highest weight and induced representations of the $\mathfrak{bms}_3$ algebra in Section \ref{sec:carroll}. We also found that the modular-T matrices in all of the aforementioned theories are essentially identity matrices, up to a phase. A natural application of the results in Section \ref{sec:carroll} is to learn about the density of primaries in Carrollian 2d CFTs in certain regimes of charges. This will be reported elsewhere \cite{Aggarwal:2025hji}. 

The techniques developed in this note are based on the mathematical properties of the known characters and are very sensitive to details. For example, our algorithm already breaks down for non-unitary warped 2d CFTs with real charge. It would be useful to develop a deeper understanding on why these methods work, either from a physical perspective, or in terms of the In\"on\"u-Wigner contraction of the Virasoro characters. One might also wonder if a  more refined mathematical algorithm exists which takes as input the characters and outputs the modular matrices for a given theory and spectrum. It would also be interesting to understand if objects like fusion and braiding matrices exist for these theories in analogy with 2d CFTs \cite{DiFrancesco:1997nk}.

We would like to finish this note by sharing our perspective on the relevance of our results for 2d Carrollian CFTs for flat holography in 3d. Our approach thus far was agnostic on this point. Recent work \cite{Cotler:2024cia,Simon:2024dwm} supports the existence of a continuous spectrum in 3d gravity. This aspect is captured by our continuous integral methods. Even if it were the case that some of the representations considered in this work may turn out to be physically irrelevant for 3d gravity, we believe our modular S-matrices should still reproduce the coarse-grained behavior that is typically captured by classical gravity. In fact, the density of primaries that one can extract from the Carrollian modular S-matrices reproduces the known Cardy formula for these theories \cite{Bagchi:2012xr,Barnich:2012xq} in the relevant regime of charges \cite{Aggarwal:2025hji}.


 \paragraph{Acknowledgments.} We thank Arjun Bagchi, Jan de Boer, St\'ephane Detournay, Daniel Grumiller and Max Riegler for collaboration on related projects and discussions. We also thank the organizers and participants of the Solvay workshop on \textit{Near-extremal black holes}. AA was supported by the Austrian Science Fund (FWF), projects P 32581, P 33789, and P 36619. JS was supported by the Science and Technology Facilities Council [grant number ST/X000494/1].

\bibliographystyle{JHEP}
\bibliography{refs}


\end{document}